# Improving Photoplethysmographic Measurements Under Motion Artifacts Using Artificial Neural Network for Personal Healthcare

Monalisa Singha Roy, Rajarshi Gupta, *Senior Member, IEEE,* Jayanta K. Chandra, *Member, IEEE*, Kaushik Das Sharma, *Senior Member, IEEE,* and Arunansu Talukdar

*Abstract*—Photoplethysmographic (PPG) measurements are susceptible to motion artifacts (MA) due to movement of the peripheral body parts. In this paper, we present a new approach to identify the MA corrupted PPG beats and then rectify the beat morphology using artificial neural network (ANN). Initially, beat quality assessment was done to identify the clean PPG beats by a pretrained feedback ANN to generate a reference beat template for each person. The PPG data were decomposed using principal component analysis (PCA) and reconstructed using fixed energy retention. A weight coefficient was assigned for each PPG sample in such a way that when they are multiplied, the modified beat morphology matches the reference template. A particle swarm optimization-based technique was utilized to select the best weight vector coefficients to tune another feedback ANN, fed with a set of significant features generated by an auto-encoder from PCA reconstructed data. For real-time implementation, this pretrained ANN was operated in feed-forward mode to directly generate the weight vectors for any subsequent measurements of PPG. The method was validated with PPG data collected from 55 human subjects. An average root-mean-square error of 0.28 and signal-to-noise ratio improvement of 14.54 dB were obtained, with an average improvement of 36% and 47% measurement accuracies on crest time and systolic to diastolic peak height ratio, respectively. With the IEEE Signal Processing Cup 2015 challenge database, Pearson's correlation coefficient between PPG-estimated and ECG-derived heart rates was 0.990. The proposed method can be useful for personal health monitoring applications.

## I. INTRODUCTION

PHOTOELECTRIC plethysmography, commonly known as photoplethysmography (PPG) or peripheral pulse signal can provide vital information on circulatory functions in human body. Although first introduced in 1937 [1], this important medical signal was underutilized for many years in cardiovascular measurements. In the context of ambulatory measurements, recent sensor deigns are focused on lightweight and in-built transmission capability [2]–[4]. Over the last decade, extensive research utilizing PPG has already established its potential for direct and indirect assessments of various cardiovascular parameters [5], [6]. An important application of PPG has been in pulse oximetry [7], which is very common in intensive care unit measurements. There have been a few attempts of utilizing the pulse transit time (distance between ECG R-peak and one PPG fiducial point) for automatic back propagation (BP) measurements [8]. Some other surrogate cardiovascular measurements such as sleep apnea classification [9], respiratory rate assessments [10], and cardiac outputs [11] have been inferred from the PPG. A model of the various cardiovascular functions such as ventricular ejection time and stiffness index from the beat-to-beat PPG morphology analysis has been proposed [12].

With the proliferation of information and communication technology in medical measurements and instrumentation, remote monitoring and analysis using handheld gadgets is a need of new generation technology. Toward this objective, a smartphone-based architecture was proposed in [13] which can utilize the in-built motion, camera, and microphone sensors for various physiological parameter monitoring, such as motion or fall, pulse oximetry, respiration, heart rate (HR), and BP. Introduction of these systems lead to personalized health care of the patients enabled with self-monitoring capability using autonomous systems. Low-power microelectronic sensors, smart textiles, and low-power wireless protocols (ZigBee, Bluetooth, etc.) are playing key roles in the development of such personal healthcare systems [14]. An important aspect of such personalized healthcare is power consumption optimization using energy harvesting technology [15].

A challenging problem with the PPG measurements has been the signal quality due to its corruption by several noise sources. This can lead to measurement inaccuracies and even wrong diagnosis. For applications involving rapid decision making, automatic quality assessment can ensure clinical acceptability of the recorded data. Research on quality assessment of PPG aims to tag or mark the recordings based on some statistical and morphological features extracted from the time-series PPG data [16]–[19]. In [20], a real-time implementation of pulse-wave contour analysis for artifact detection in PPG is presented.

The existing research on MA reduction from PPG can be categorized into two broad categories, viz., adaptive digital signal processing and statistical signal processing. Some of the reported works used a separate motion sensor (accelerometer) so that the reconstructed data can be compared with the clean PPG signal. Among the initial signal processing tools were independent component analysis (ICA) [21], [30], Fourier series analysis [22]–[24], principal component analysis (PCA) [23], [26], fixed interval Kalman smoother [24], and adaptive noise canceller [27]. Krishnan *et al.* [23] propose a motion detection unit which assessed the extent of corruption by skew, kurtosis, and bispectrum and then used a Fourier series reconstructor for MA reduction. A real-time implementation of MA reduction is described in [25] using a ratiometric technique, which relies on a model that motion-induced absorption in PPG sensor to be multiplicative and independent of capillary blood volume. In [27], a synthetic representation of the noise model is generated from the MA corrupted PPG data, thus, eliminating the requirement of any additional accelerometer. A modified ICA-based dual-tree discrete wavelet transform method is described in [28]. Moreover, the method also retains the respiration information in the restored PPG. Motion tolerant PPG wearable applications are described in [29], [34], and [35]. Among these, an average beat template was computed and matched in [34]. A time-domain ICA (TD-ICA) approach for iterative motion reduction based on singular spectral analysis using PPG and accelerometer data is reported in [31]. A spectral filtering algorithm in which time varying power spectral density is calculated for both PPG and accelerometer data to detect the motion artifacts (MA) and successfully reconstructed the corrupted signal is described in [32]. Peng *et al.* [33] describe a wavelet decomposition combined with comb filtering approach applied for MA reduction in PPG. In some works, a synthetic noise model has been developed either from physiological aspects (tissue, venous blood movements, etc.) or, extracting motion information from corrupted PPG. A minimum noise estimate filter for electrocardiogram and electroencephalogram is described in [36].

The motivation of this paper is to develop a new PPG preprocessing method, which is independent of the MA characteristic. In real life the noise behavior may change and, a fixed noise model developed from an acquired data segment may not be effective for continuous measurements. In this paper, we propose a new approach for MA reduction from the PPG signal in personal healthcare applications. It is based on the hypothesis that each human individual has a specific pulse wave signature. Although this signature slowly varies over time due to age and various cardiovascular factors, its variations over short period of time are minor, unless there is sudden major cardiovascular abnormality. We propose a two-stage approach, viz., to assess the quality of the individual beats through a neural network (NN) classifier, followed by an optimization technique to refine the morphology of the corrupted beats based on a reference template beat generated from the same subject. There are two specific contributions in the proposed technique.

1) It deals with the quality assessment of individual PPG beats instead of the whole time-series data, mostly dealt in other published research except [34].
2) It does not require either a separate accelerometer or a clean PPG, and thus, fully adaptive in nature.

The rest of this paper layout is as follows. Section II elaborates the beat classification, quality assessment, and artificial neural network (ANN)-based PPG preprocessing technique. Section III describes various objective and subjective performance assessment parameters using normal and cardiovascular subjects. Section IV summarizes the main outcome of this paper.

## II. METHODOLOGY

Processing stages of the proposed technique, as shown in Fig. 1 and detailed in Sections II-A–II-F, consist of two main process flow paths. The first, PPG beat quality evaluation using a multilayer perceptron-neural network (MLP-NN) classifier and generating the reference beat template for each subject and updating of respective global template. The MLP-NN classifier was trained by clean and noisy PPG beats. The second is PPG preprocessing using a combination of a deep auto-encoder (DAE) and MLP-NN-based MA reduction model. During the training phase, particle swarm optimization (PSO) has been employed to train the MA reduction model, fed with reduced seat of features from a PCA decomposed PPG beat matrix. In the testing stage, the PSO-based evaluation of weight matrix will not be performed. As the proposed scheme is person specific, the MLP-NN-based MA reduction model requires periodic recalibration using the PSO. The training and testing paths of these NNs are shown separately.

### A. PPG Data Collection Protocol

Total 30 healthy and 25 patients with cardiovascular disease (CVD) were engaged in the study for six months. Informed consent was taken from each volunteer after explaining the purpose of our study. From each subject, 5-min data were collected using a transmission-type PPG sensor [37] attached to the left hand, kept stationary. Simultaneously, the motion data were collected by an accelerometer attached to right hand moved horizontally (elbow movement). A data acquisition module (USB 6009 from National Instruments) was used to collect PPG and motion data at 14-bit resolution and at 60-Hz sampling in a computer. The noisy PPG signal was synthetically generated by sample to sample addition of clean PPG and motion data.

### B. Beat Detection, Quality Assessment of PPG, and Formation of a Reference Template

At the first stage, beat detection and delineation was performed from the PPG data array using the acceleration PPG array and running amplitude comparison between two consecutive samples with thresholding, as described in [38].

The average beat length was computed separately for anacrotic phase and catacrotic phase of individual beats. Then, keeping the relative position (index) of systolic peaks same in all beats, the length toward beginning or/and tail of each beat were truncated/padded at both beat ends to match this average HR by truncation and extrapolation method [39]. The HR information of the individual beats was separately stored. A beat matrix $[B]_{n \times p}$ was formed as

$$B = [\text{beat}_1 \ \text{beat}_2 \ \ldots \ \text{beat}_p]_{n \times p}$$
$$\text{beat}_k = \begin{bmatrix} x_1^k & x_2^k & x_3^k & \ldots & x_n^k \end{bmatrix}^T \quad (1)$$

where $p$ is number of beats and $n$ represents the number of samples in each beat. In the next stage, the quality assessment of the individual beats was performed using an MLP-NN classifier with six hidden layers. The whole PPG beat was fed as input and a binary classifier was designed. As the proposed scheme is person specific, the ANN was trained by a back-propagation algorithm with a set of 120 clean and 120 noisy PPG beats for each subject. Six fold cross validation and drop-out technique were used to avoid the over/under fitting of the NN. The "clean" beats are marked which have high correlation with the global reference beat template. Hence, if $g$: number of acceptable beats out of $p$, then by averaging the matrix $[h]_{n \times g}$ a target beat was formed as $[t]_{n \times 1}$. This was used as the reference template for current measurement and also to update the global reference beat template for future measurements and recalibration purpose, if necessary.

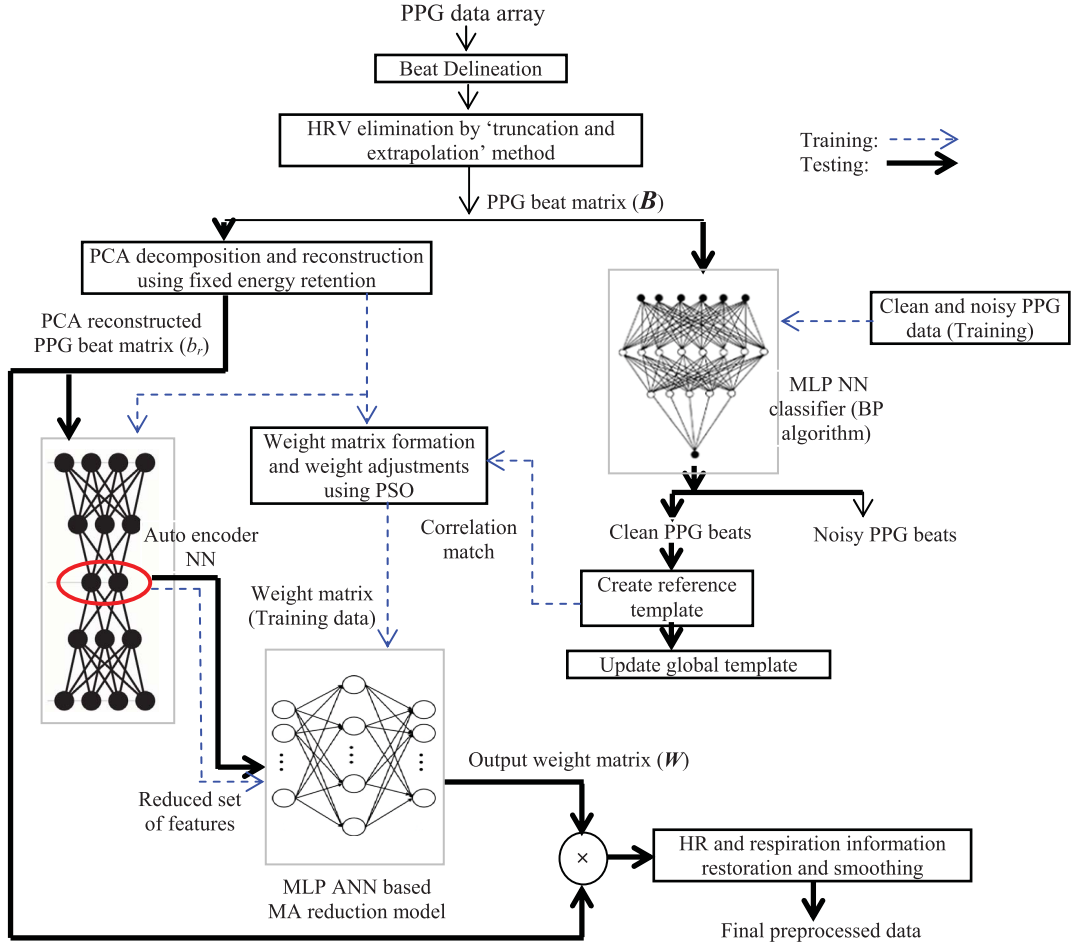

Fig. 1. Preprocessing block diagram.

### C. Principal Component Analysis (PCA) of the Beat Matrix

The beat matrix $B$ was decomposed using a PCA obtained by eigenvalue decomposition of its covariance matrix. The linear orthogonal expansion is represented as

$$[\text{PC}]_{p \times n} = \Psi^T \times [B]$$
$$\text{PC} = [\text{pc}_1 \ \text{pc}_2 \ldots \ \text{pc}_p]^T \quad (2)$$

where PC is the principal component matrix. $\Psi$ is a $p \times p$ eigenvector matrix of $B^T B$. These PCs were arranged as decreasing number of variances and could provide their contributions (in terms of significant or insignificant) in the original PPG data. Since the collective contributions from the lower order PCs are very low, the original beat matrix can be reconstructed using less number of PCs with some error. In the proposed scheme the beat matrix was reconstructed using fixed energy retention (of 95%). Energy contribution from the $k$th PC was calculated as follows:

$$E_k = \frac{\sum_{j=1}^{n} y_j^2}{\sum_{k=1}^{p} \left[ \sum_{j=1}^{n} y_j^2 \right]} \quad (3)$$

where $y$s are the PC elements. The remaining PCs were made zero along with the corresponding eigenvectors, and the signal can now be reconstructed using the formula

$$[b_r]_{p\times n} = \Psi_r \times [PC_r]_{p\times n} \qquad (4)$$

where $\boldsymbol{b_r}$: reconstructed beat matrix; $\boldsymbol{\Psi_r}$: modified eigenvector matrix, and $\boldsymbol{PC_r}$: modified principal component matrix.

### D. Weight Matrix Formation

The reconstructed beat matrix now can be represented as

$$b_r = W \circ B = \begin{pmatrix} w_1^1 & w_1^2 & \dots & w_1^p \\ w_2^1 & w_2^2 & \dots & w_2^p \\ & \dots & & \\ w_n^1 & w_n^2 & \dots & w_n^p \end{pmatrix} \circ \begin{bmatrix} x_1^1 & x_1^2 & \dots & x_1^p \\ x_2^1 & x_2^2 & \dots & x_2^p \\ \dots & \dots & \dots & \dots \\ x_n^1 & x_n^2 & \dots & x_n^p \end{bmatrix} \qquad (5)$$

where $\boldsymbol{W}$: the weight matrix, and "$\circ$" represents Hadamard product of $\boldsymbol{W}$ and $\boldsymbol{B}$. Thus, morphology of $k$th beat may be refined to match the reference template by adjustment of the corresponding row-weight vector. The proposed scheme utilized PSO, as a stochastic optimization technique, to adjust each of the row-weight vector coefficients $(w_j^k)$ based on correlation match between the $k$th beat and reference template. The fitness function (ff) to evaluate the optimal weight matrix ($\boldsymbol{W}$) employing PSO, was formulated as

$$\text{ff} = \text{corr}(b_r^k, T), \quad \text{for} \quad \forall k \qquad (6)$$

where $b_r^k$ is the $k$th beat of $\boldsymbol{b_r}$, and $corr$: correlation coefficient between the current beat and reference template $(T)$. The stopping criteria of PSO algorithm was set as the achievement of 99% correlation between $T$ and $b_r^k$. The set of optimally tuned weight vectors for each subjects act as the training data for the next stage of processing.

### E. Feature Extraction Using Deep Auto-Encoder for ANN Training

PSO being an offline technique cannot be used for real-time implementation. Hence, for generation of $\boldsymbol{W}$ for real-time measurements, an MLP-NN MA reduction model was used. This ANN was fed by a reduced set of features, tapped from a six-layer DAE. A special form of ANN, DAE [40] can extract effective features from a high-dimensional data in an unsupervised manner. The basic organization of a DAE resembles a funnel shaped structure, with few intermediate layers. The first few layers have gradually decreasing number of nodes, performing the encoder function, followed by gradually increasing number of nodes, performing the decoding task. In this proposed person specific PPG MA reduction technique, the DAE was fed with PCA reprocessed PPG beat samples ($b_r^k$), which mapped the $n$ number of preprocessed samples ($\cong$38–45) to 12 features at the third hidden layer. Its objective is to train the following function:

$$b_r^k \approx c^k \qquad (7)$$

where $c^k$ is the output of the encoder corresponding to $(b_r^k)$. The encoder function and decoder function was represented as

$$\begin{aligned} h_1^k &= f(E_1^k \times b_r^k + q_1^k) \quad \forall k, \text{ for first hidden layer} \\ c^k|_j &= f(E_j^k \times h_j^k + q_j^k) \quad \forall k, \text{ for } j\text{th hidden layer} \end{aligned} \qquad (8)$$

where $E_j^k$ denotes the unit connection between layer $j$ and $j+1$ for the $k$th beat, $h_j^k$ is the $j$th hidden layer, and $q_j^k$ denotes the bias term associated with unit in layer $j$, $c^k|_j$ the output from $j$th hidden layer corresponding to $k$th beat, and $f(\cdot)$ denotes the activation function. $h_j^k$ represents the feature vector or compressed representation of the input data set $(b_r^k)$. The objective now becomes to minimize the function

$$J(E_{j+1}^k, q_j^k, q_{j+1}^k) = \min \sum [b_r^k - c^k]_2^2. \qquad (9)$$

In this paper, the input $(b_r^k)$ to the auto-encoder is $n$-dimensional and output is a set of $m$ features where, $m < n$. The objective function $J$ was minimized employing BP algorithm. During the training stage the output $c^k$ of the auto-encoder was testified with its input $b_r^k$ and that intern guaranteed that the features extracted from the intermediate third hidden layer was mature enough to construct the MLP NN-based MA reduction model. In the training stage the NN was used in BP mode, utilizing the PSO generated weight vector as reference (training data). The NN had $m$ (=12) input nodes, six hidden layers and the number of output equals number of weight coefficients in $\boldsymbol{W}$. Drop out technique was utilized to avoid over/under fitting of the MA reduction model. For an $n \times p$-dimensional beat matrix, the pretrained ANN adjusts its weights in such a way so that the magnitude of $p$-dimensional error goes below a small predefined threshold value. Finally, in real-time measurements, this MLP-NN MA reduction model operates in feed-forward mode. This accepts the beat matrix $\boldsymbol{b_r}$ and directly generates the weight matrix $\boldsymbol{W}$ using (5).

### F. Heart Rate Information Retrieval in Data

At this final stage, the HR information of the individual beats was restored. Since the beat ends may not match exactly to the same of neighboring ones, intermediate PPG samples were generated by using an adaptive spline interpolation technique.

## III. RESULTS AND DISCUSSION

In the first stage, we describe the beat detection and classification results from PPG data using 55 numbers of volunteers' data, involving 13 100 numbers of PPG beats. The classification performance parameters are defined as

$$\begin{aligned} \text{Sensitivity (SE)} &= \frac{\text{TP}}{\text{TP} + \text{FN}} \times 100 \\ \text{Positive Predictivity (PP)} &= \frac{\text{TP}}{\text{TP} + \text{FP}} \times 100 \end{aligned} \qquad (10)$$

where TP (true-positive): correctly detected noisy beats, FN (false negative): correctly detected clean beats, and FP (false positive): misdetected noisy beats.

In Fig. 2(a)–(c), the classification results [SE, PP, and area under the curve (AUC)] are shown separately for 10 subjects,

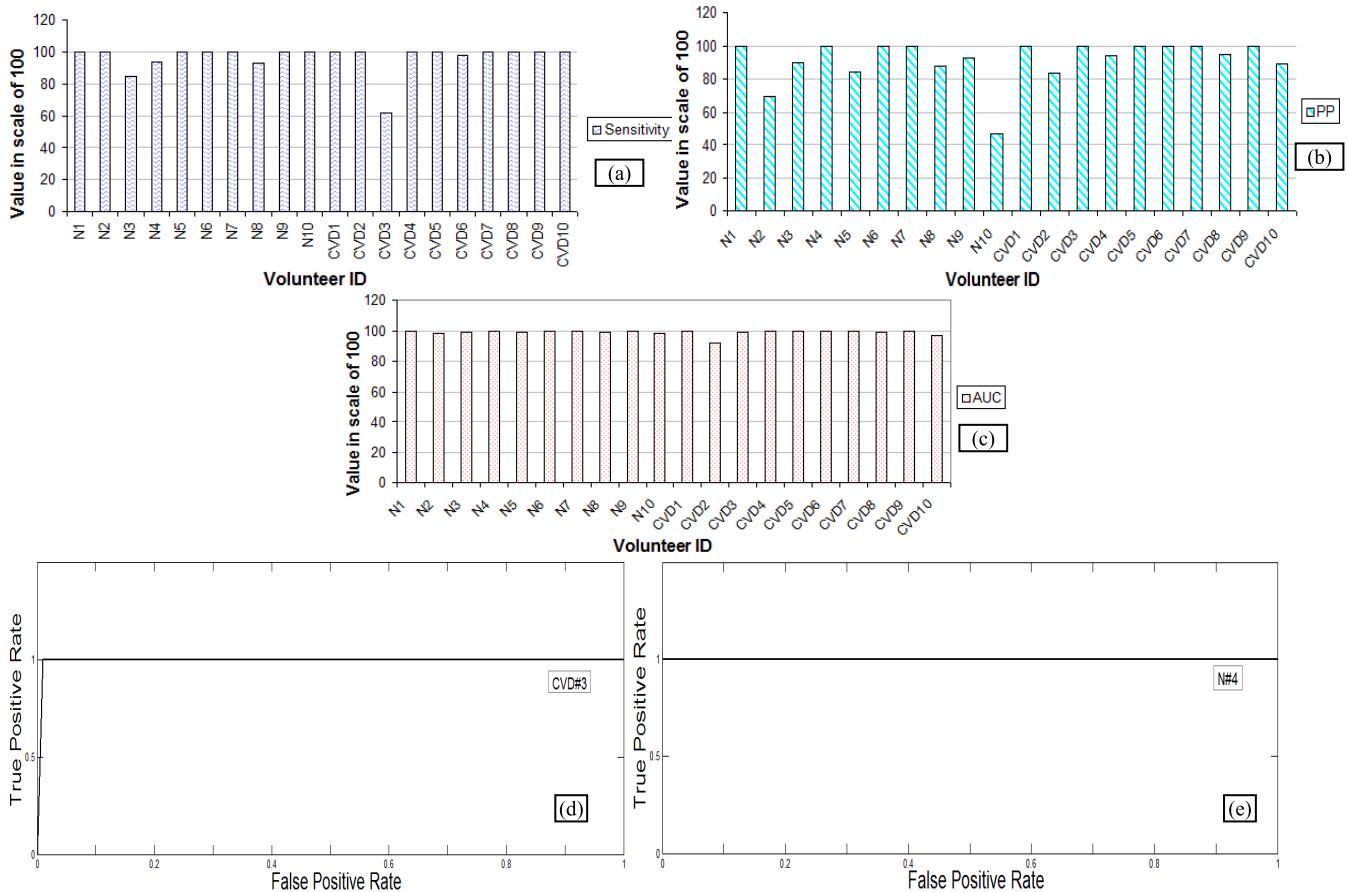

Fig. 2. PPG beat classification results with N and CVD groups. (a)–(c) Subject wise. (d) ROC curve for CVD#3. (e) ROC curve for N#4.

taken from each group (healthy and CVD). For each subject, 120 noisy and clean beats were used for training the NN classifier. Collectively, for 30 N group (CVD group), the average SE and PP were found as 98.12% (97.95%) and 92.24% (98.66%), respectively. The receiver operating characteristics (ROC) curves for N#4 and CVD#3 are shown in Fig. 2(d) and (e), respectively. The AUC values for N and CVD groups were found as 0.99 and 0.98, respectively. The AUC values (in a scale of 100) for the 20 subjects are also shown in Fig. 2(a). Fig. 3 shows the beat-wise performance of the proposed PCA-PSO preprocessing algorithm. Fig. 3(a) shows the "reference beat" with one clean beat and arbitrarily selected four corrupted beats, all selected from N#4 data. The initial correlation coefficients with the reference beat are also indicated. Fig. 3(b) shows the same beats preprocessed through the PCA and PSO, with the improved correlation coefficients indicated. The clean beat, having maximum morphological similarity with the reference beat has an improvement of 0.01 only. Fig. 4 shows the beat-wise comparative performance in preprocessing between PSO and ANN methods using arbitrarily PPG beats from N#4 data. In each panel, the reference beat, initial corrupted beat (after HR elimination), and the final preprocessed beat by PCA-PSO and ANN are shown in separate colors. The correlation coefficients between the preprocessed beat and reference beat are indicated. It is observed that both methods nearly achieve the same final correlation. The qualitative reconstruction of the data is shown in Fig. 5(a) and (b) for two volunteers, one healthy and other a CVD patient. In each illustration, the top panel shows the clean signal, with the noisy signal in the middle panel.

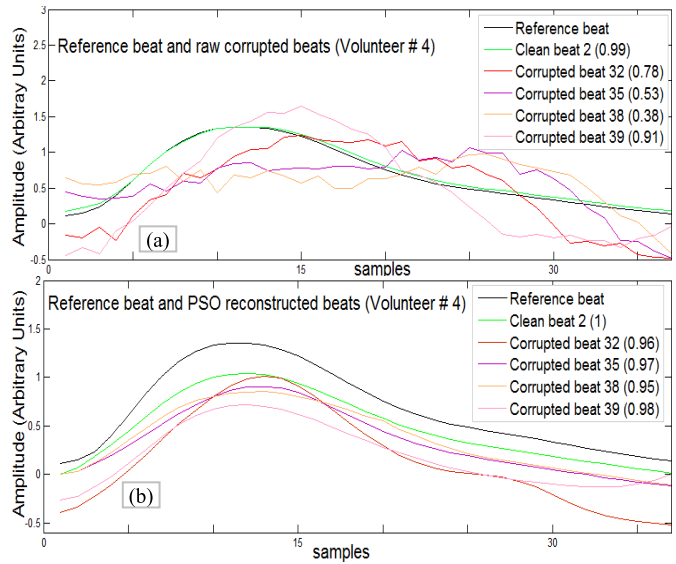

Fig. 3. Beat-wise preprocessing performance using PCA and PSO for N#4. (a) For raw corrupted beats. (b) After PSO preprocessing.

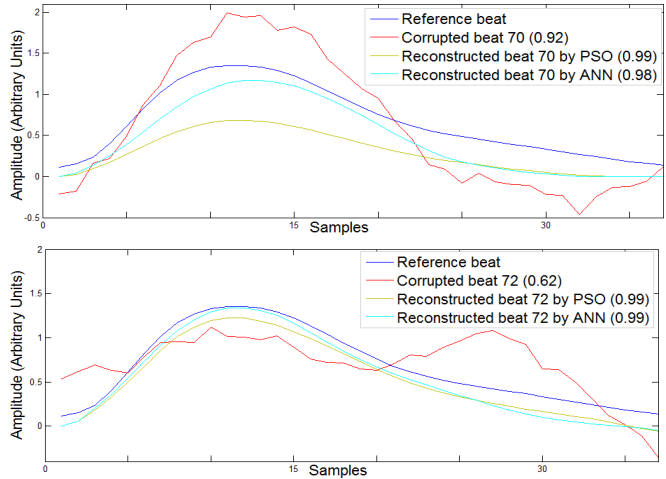

Fig. 4. Comparison of preprocessing performance between PSO and ANN for $N\#4$ for two separate beats.

The PCA-PSO and ANN preprocessed data are shown in lower two panels. In Fig. 5(a), beats 32–34 and 40–43 are heavily corrupted and marked. The root-mean-square error (RMSE) value between the original and preprocessed beats obtained by PCA-PSO and ANN were found as 0.20 and 0.18, respectively. The RMSE between weight matrixes obtained by PCA-PSO and ANN was 0.12. For Fig. 5(b), the beats 15–17, 24–26, and 32–47 are heavily corrupted and rest is comparatively cleaner. The RMSEs between the original and preprocessed data obtained by PCA-PSO and ANN were found as 0.13 and 0.13, respectively. The RMSE between weight matrixes obtained by PCA-PSO and ANN was 0.22. For both data, the PSO and ANN methods could satisfactorily denoise all these corrupted regions. All the preprocessed data were clinically verified and acceptable for further clinical evaluation. For evaluating denoising performance, the objective assessment was done by computing RMSE and signal-to-noise ratio (SNR) improvement

$$\text{RMSE} = \sqrt{\frac{1}{N}\sum_i [x(i)-\tilde{x}(i)]^2}$$

$$\text{SNR improvement (SNR}_i) = 10 \times \log \frac{\sum_i \frac{x^2(i)}{n^2(i)}}{\sum_i \frac{x^2(i)}{[x(i)-\tilde{x}(i)]^2}} \quad (11)$$

where $x$ = clean signal, $\tilde{x}$ = reconstructed signal, and $n$: residual noise.

The error metrics for evaluating the preprocessing performance for arbitrarily chosen five volunteers each from $N$ and CVD groups are shown in Table I. The average RMSE and SNR$_i$ computed over $N$ group (CVD group) were found as 0.19 (0.31) and 13.99 dB (14.45 dB), respectively. It also shows the correlation coefficient (CORR) initial and final values. For CVD#3, highest correlation coefficient improvement value was achieved, from 0.49 to 0.90.

To assess the noise sensitivity of the proposed technique, we introduced various noise levels (in dBs) in different segments of the same data set $N$. Noise levels of 10, 15, and 30 dB were synthetically added in three consecutive segments, each of length 700 samples in $N\#4$ data, as shown in Fig. 6(a). The noisy data sets to these noise levels are shown at left and corresponding preprocessed data at right in the next three consecutive panels [Fig. 6(b)–(d)]. The RMSE, SNR$_i$ for segment 1 were 0.20, and 24.22, respectively, for segment 2 were 0.21 and 17.85, respectively, and for segment 3 were 0.24 and 5.63, respectively. It can be concluded that the proposed algorithm performs equally well for different noise levels. To justify the clinical acceptance of the preprocessed data, number of acceptable PPG beats those could be utilized for diagnostic measurements were compared with the same from noisy data. Two important clinical features, viz., crest time, and systolic to diastolic peak height ratio (a/b) were considered.

TABLE I
PERFORMANCE PARAMETERS OF PREPROCESSING

| Volunteer ID [Age, M/F, Height(in cm), Weight (in kg), Group] | RMSE | SNR$_i$ | CORR (initial) | CORR (final) |
|---|---|---|---|---|
| N#1 23,M,164,60) | 0.27 | 25.90 | 0.78 | 0.96 |
| N#2 (25,M,170,68) | 0.21 | 9.01 | 0.87 | 0.94 |
| N#3 (21,M,160,55) | 0.32 | 4.38 | 0.67 | 0.79 |
| N#4 (23,M,169,70) | 0.20 | 19.21 | 0.77 | 0.94 |
| N#5 (23,M,172,69) | 0.12 | 6.70 | 0.88 | 0.93 |
| CVD#1 45,F,152,72) | 0.68 | 1.58 | 0.78 | 0.95 |
| CVD#2 41,F,161,70) | 0.38 | 19.18 | 0.91 | 0.98 |
| CVD#3 (70,M,172,68) | 0.13 | 24.17 | 0.49 | 0.90 |
| CVD#4 (51,F,163,72) | 0.35 | 19.28 | 0.88 | 0.96 |
| CVD#5 (59,M,175,61) | 0.10 | 11.48 | 0.69 | 0.91 |
| Notations: N: Normal; CVD: cardiovascular patient | | | | |

The percentage improvements were calculated as

$$\text{Improvement} = \frac{N_p - N_n}{N_p} \times 100 \quad (12)$$

where $N_p(N_n)$ represents the number of beats corresponding to preprocessed (noisy) data.

Table II shows the figures for the same set volunteers' data. An average improvement in crest time and a/b measurement acceptability for $N$ group (CVD group) were found as 38.2% (33.01%) and 50.22% (42.41%), respectively. This means that, the technique could improve the clinical acceptability of more than one-third corrupted beats for crest time measurement and nearly half number of corrupted beats for a/b measurement, both for $N$ and CVD groups.

The performance comparison with other published works is given in Table III. For this, we used motion corrupted PPG data from the IEEE Signal Processing cup 2015 [41]. Total six channels: single lead chest ECG, two-channel PPG from wrist, and three-axis acceleration was collected from 24 human subjects who are put at intense physical exercise. The challenge is to calculate the HR from the preprocessed PPG and compare with that extracted from ECG, taken as the

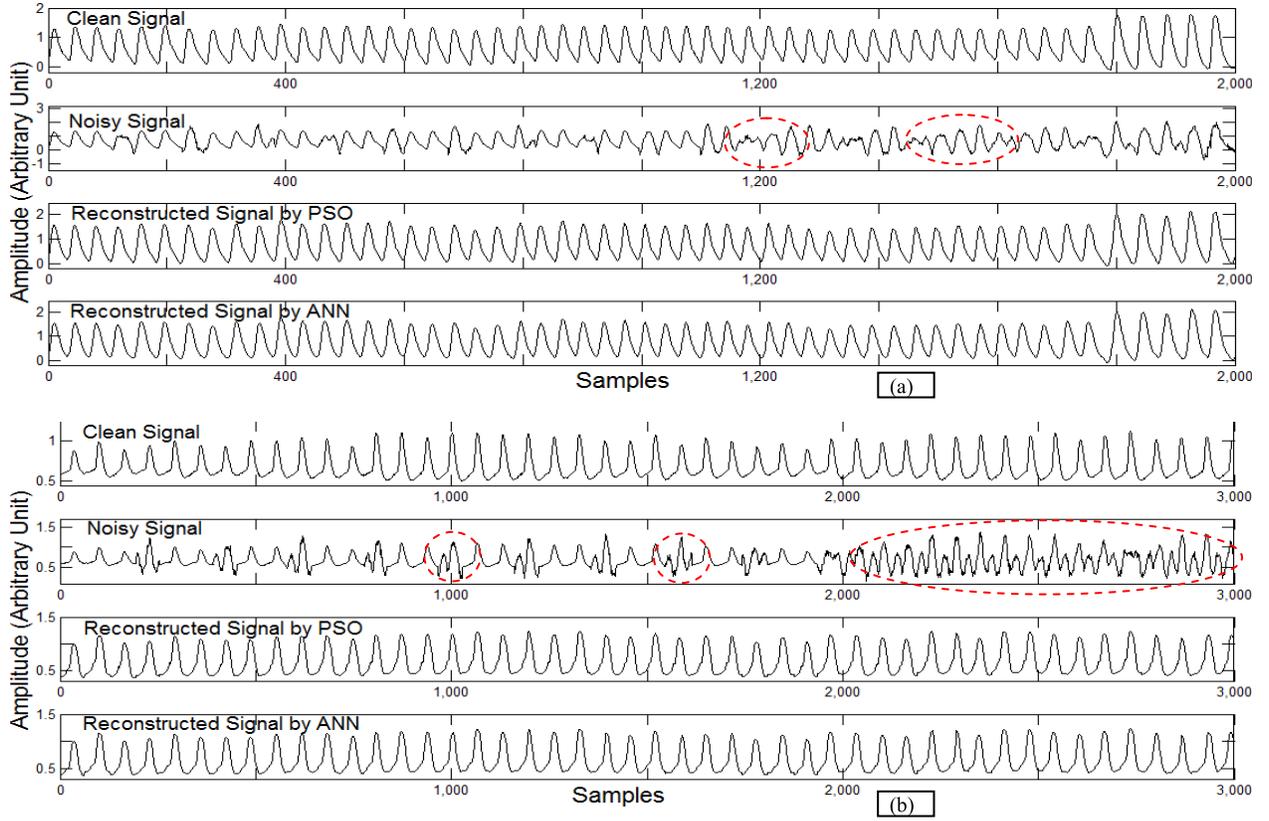

Fig. 5. Preprocessing results for the subjects. (a) *N*#4 (normal) and (b) CVD#20.

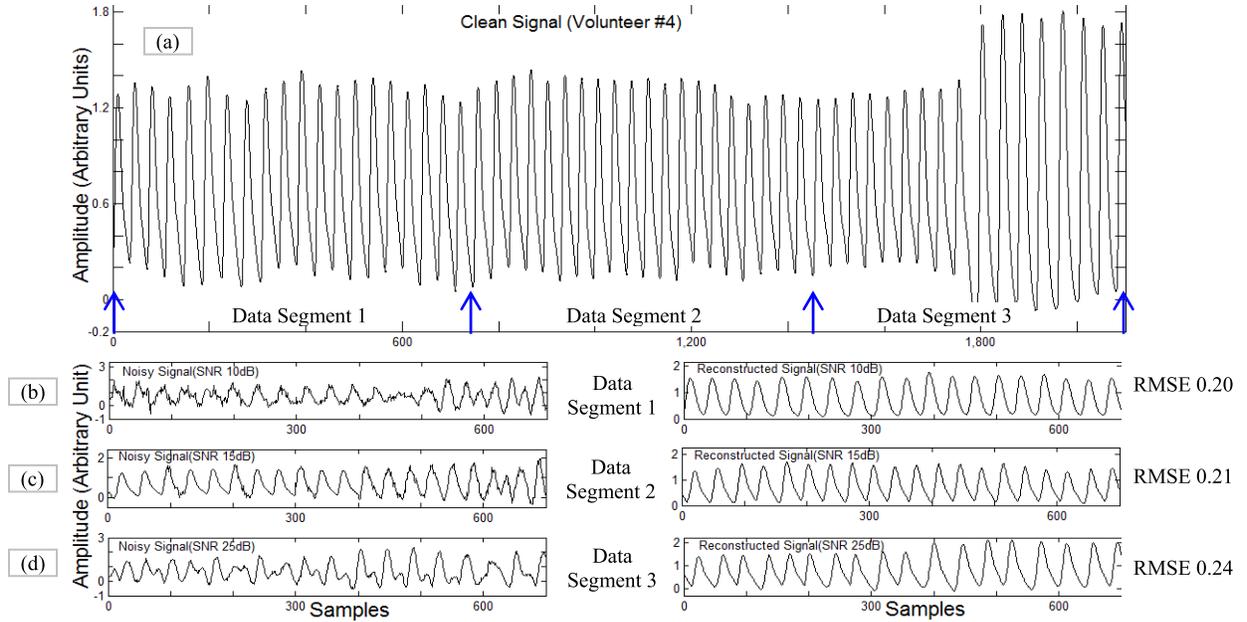

Fig. 6. Noise sensitivity of the proposed algorithm for the subject *N*#4. Performance with (a) 10 dB, (b) 15 dB, and (c) 25 dB noise level.

"ground truth." Two error parameters are defined as

$$\text{Error1} = \frac{1}{N}\sum_{i=1}^{N}|\text{BPM}_{\text{est}}(i) - \text{BPM}_t(i)|$$

$$\text{Error2} = \frac{1}{N}\sum_{i=1}^{N}\frac{\sum_{i=1}^{N}|\text{BPM}_{\text{est}}(i) - \text{BPM}_t(i)|}{\text{BPM}_t(i)} \quad (13)$$

where $N$ is the total number of windows, $\text{BPM}_{\text{est}}$ is the estimated HR using preprocessed PPG, and $\text{BPM}_t$ is the true HR (ground truth) in the same window. While Error1 proves the average error in HR estimation, Error2 provides average percentage error of estimated HR. Table III shows the comparison of error parameters and Pearson's correlation coefficient with other published works using the same database. It shows that our results are comparable to that of the published works.

TABLE II
IMPROVEMENT OF WAVE DIAGNOSIS FROM RECONSTRUCTED PPG

| Volunteer ID [Age, M/F, Height(in cm), Weight (in kg), Group] | Improvement in crest time measurements (in percentage) | Improvement in a/b ratio measurements (in percentage) |
|---|---|---|
| N#1 (23,M,164,60) | 15.49 | 64.78 |
| N#2 (25,M,170,68) | 31.25 | 39.58 |
| N#3 (21,M,160,55) | 70.14 | 59.70 |
| N#4 (23,M,169,70) | 36.25 | 37.50 |
| N#5 (23,M,172,69) | 33.33 | 46.96 |
| CVD#1 (45,F,152,72) | 14.49 | 30.28 |
| CVD#2 (41,F,161,70) | 16.6 | 29.61 |
| CVD#3 (70,M,172,68) | 25.41 | 37.18 |
| CVD#4 (51,F,163,72) | 65.21 | 54.35 |
| CVD#5 (59,M,175,61) | 39.68 | 55.90 |
| Notations: N: Normal; CVD: cardiovascular patient | | |

TABLE III
PERFORMANCE COMPARISON WITH OTHER WORKS

| Published Work | Error 1 | Error 2 (%) | Correlation Coefficient (CC) |
|---|---|---|---|
| S.M.A. Salehizadeh, et al. [32] | 0.91 | 0.67 | 0.990 |
| Y. Ye, et al. [35] | 1.16 | 0.90 | 0.995 |
| Z. Zhang et al. [41] | 2.35 | 1.76 | 0.992 |
| B. Sun et al. [42] | 1.41 | 1.12 | 0.995 |
| Z. Zhang [43] | 1.29 | 1.09 | 0.993 |
| Y. Ye, et al. [44] | 1.22 | - | 0.992 |
| Proposed method | 1.47 | 1.10 | 0.990 |

Although this paper is aimed for personalized healthcare, the MLP ANN-based MA reduction model needs periodic calibration. This is ascertained by a cross correlation (CC) between the current reference template and the global template of the patient (below a CC level of 0.7). In such a case, the training of DAE and MLP NN model should be reperformed with the PSO.

The average time complexity of the ANN algorithm was found as 625 ms per beat Machine Configuration: 16-GB RAM, AMD processor, 3.1-GHz clock speed, and 1600-MHz RAM speed. The success of the proposed method depends upon two important factors. The MA corrupted PPG data should have a few (30% of total) clinically acceptable beats. If it is fully corrupted with MA, then the technique works satisfactorily as long as the global reference template successfully represents the morphology of the current PPG beats. Second, the trained ANN weight vector pattern is unique for an individual and not universal for all. Second, when the correlation between the reference template beat and the global template falls below 0.75, a new calibration is suggested. The calibration process involves retraining of the MLP-NN-based MA reduction model using fresh data.

## IV. CONCLUSION

This paper describes a new preprocessing technique based on ANN for finger PPG signal corrupted with MA. The proposed technique eliminates the requirement of additional hardware and needs only a reference beat template of the same subject to be prestored in system memory. A salient contribution of this paper is that the reference template is extracted from the acceptable quality beats of the current data through a beat quality assessment technique. For beat classification, we found an average accuracy of 98.03% over nearly 13 100 number of PPG beats. The technique performs satisfactorily as long as the global reference template represents the clean beat morphology of the subject. The proposed technique could offer an improvement of 35.92% and 47.22% for crest time and systolic to diastolic peak amplitude ratio ($a/b$) measurement, respectively, from corrupted PPG data. It provided equally good performance with different noise levels. The average SNR improvement of 14.54 dB and RMSE of 0.28 were found. It is also noteworthy that our results are comparable with published works in PPG preprocessing literature.

ACKNOWLEDGMENT

The authors would like to thank the ongoing SAP DRS II Program from 2015 to 2020, from the University Grants Commission, India, at the Department of Applied Physics, University of Calcutta, Kolkata, India. They would also like to thank the reviewers for constructive criticism which helped to enhance the readability of the manuscript. M. S. Roy would like to thank UPE-II Project at the University of Calcutta.

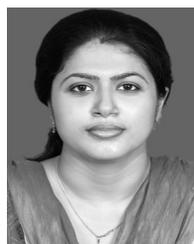

**Monalisa Singha Roy** received the B.Tech. degree in electronics and instrumentation engineering from the West Bengal University of Technology, Kolkata, India, in 2007, and the M.Tech. degree in instrumentation and control engineering from the Department of Applied Physics, University of Calcutta, Kolkata, in 2011.

She has six years of industrial research and development experience in embedded system in power electronics. She is currently a Junior Research Fellow with the Department of Applied Physics, University of Calcutta, under the University Potential for Excellence-II Project from the University Grants Commission, India. Her current research interests include cardiovascular signal processing.

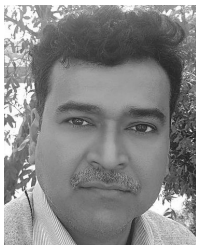

**Rajarshi Gupta** (M'07–SM'14) received the B.Tech., M.Tech., and Ph.D. degrees in instrumentation engineering from the University of Calcutta, Kolkata, India, in 1998, 2002, and 2012, respectively.

He is currently an Associate Professor with the Electrical Engineering Section, Department of Applied Physics, University of Calcutta. He is engaged as an investigator in four research projects involving development of biomedical applications and systems. He has authored 45 publications in international journals and conferences. His current research interests include biomedical signal analysis and compression, and intelligent health monitoring systems.

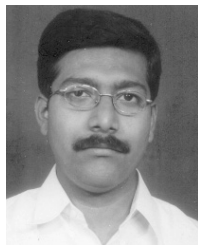

**Kaushik Das Sharma** (M'12–SM'16) received the B.Tech. and M.Tech. degrees in electrical engineering from the Department of Applied Physics, University of Calcutta, Kolkata, India, in 2001 and 2004, respectively, and the Ph.D. degree from Jadavpur University, Kolkata, in 2012.

He is currently an Associate Professor with the Electrical Engineering Section, Department of Applied Physics, University of Calcutta. He has authored or co-authored about 45 technical articles, including 20 international journal papers. His current research interests include fuzzy control, stochastic optimization, machine learning, robotics, and systems biology.

Dr. Das Sharma has served/is serving in important positions in several international conference committees all over the world. He also serves as the Secretary of the IEEE JOINT CSS-IMS KOLKATA CHAPTER.

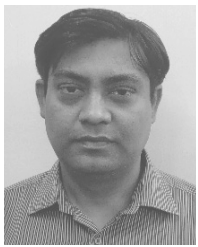

**Jayanta K. Chandra** (M'12) received the B.Tech., M.Tech., and Ph.D. degrees in electrical engineering from the Department of Applied Physics, University of Calcutta, Kolkata, India, in 2001, 2004, and 2014, respectively.

He is currently an Associate Professor with the Department of Electrical Engineering, Purulia Government Engineering College, Purulia, India. He has authored or co-authored about 20 technical articles, including six international journal papers. His current research interests include machine learning and image processing.

Dr. Chandra has served/is serving in important positions in several international conference committees in India.

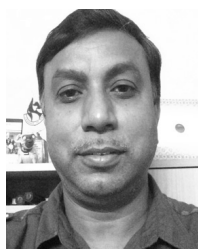

**Arunansu Talukdar** received the M.B.B.S. and M.D. (Medicine) degrees from the University of Calcutta, Kolkata, India, and the Ph.D. degree from the University of California at Los Angeles, Los Angeles, CA, USA.

He is currently a Professor of Medicine, Medical College, Kolkata, India, where he is also a Nodal Officer of the Multidisciplinary Research Unit. He has authored over 45 original articles in different national/international journals of repute and many book chapters. His current research interests include biomedical device developments.

Dr. Talukdar is an Expert Committee Member of the Drug Controller General of India. He is a fellow of the World Health Organization and Indian College of Physicians. He was a recipient of the Fogarty International Scholarship.